\documentclass[reprint, superscriptaddress, twocolumn, amsmath, amssymb, aps, prb]{revtex4-2}
\usepackage{graphicx}
\usepackage{dcolumn}
\usepackage{bm}
\usepackage{placeins}
\usepackage{appendix}
\usepackage{upgreek}

\usepackage{verbatim}
\usepackage[usenames,dvipsnames]{xcolor}
 \usepackage{amsmath}
 \usepackage{amsfonts}
 \usepackage{amssymb}
 \usepackage[colorlinks=true,citecolor=blue,linkcolor=red]{hyperref}

 \usepackage{bbold}     
 \usepackage[makeroom]{cancel}  
 \usepackage{multirow}    
 \usepackage[normalem]{ulem}        
 \usepackage{array}
 \usepackage{pdfpages}
\makeatletter
 \AtBeginDocument{\let\LS@rot\@undefined}
 \makeatother

\begin{document}

\title{SQUID oscillations in PbTe nanowire networks}

\author{Yichun Gao}
\email{equal contribution}
\affiliation{State Key Laboratory of Low Dimensional Quantum Physics, Department of Physics, Tsinghua University, Beijing 100084, China}

\author{Wenyu Song}
\email{equal contribution}
\affiliation{State Key Laboratory of Low Dimensional Quantum Physics, Department of Physics, Tsinghua University, Beijing 100084, China}

\author{Zehao Yu}
\email{equal contribution}
\affiliation{State Key Laboratory of Low Dimensional Quantum Physics, Department of Physics, Tsinghua University, Beijing 100084, China}

\author{Shuai Yang}
\affiliation{State Key Laboratory of Low Dimensional Quantum Physics, Department of Physics, Tsinghua University, Beijing 100084, China}

\author{Yuhao Wang}
\affiliation{State Key Laboratory of Low Dimensional Quantum Physics, Department of Physics, Tsinghua University, Beijing 100084, China}

\author{Ruidong Li}
\affiliation{State Key Laboratory of Low Dimensional Quantum Physics, Department of Physics, Tsinghua University, Beijing 100084, China}

\author{Fangting Chen}
\affiliation{State Key Laboratory of Low Dimensional Quantum Physics, Department of Physics, Tsinghua University, Beijing 100084, China}

\author{Zuhan Geng}
\affiliation{State Key Laboratory of Low Dimensional Quantum Physics, Department of Physics, Tsinghua University, Beijing 100084, China}

\author{Lining Yang}
\affiliation{State Key Laboratory of Low Dimensional Quantum Physics, Department of Physics, Tsinghua University, Beijing 100084, China}

\author{Jiaye Xu}
\affiliation{State Key Laboratory of Low Dimensional Quantum Physics, Department of Physics, Tsinghua University, Beijing 100084, China}

\author{Zhaoyu Wang}
\affiliation{State Key Laboratory of Low Dimensional Quantum Physics, Department of Physics, Tsinghua University, Beijing 100084, China}

\author{Zonglin Li}
\affiliation{State Key Laboratory of Low Dimensional Quantum Physics, Department of Physics, Tsinghua University, Beijing 100084, China}

\author{Shan Zhang}
\affiliation{State Key Laboratory of Low Dimensional Quantum Physics, Department of Physics, Tsinghua University, Beijing 100084, China}

\author{Xiao Feng}
\affiliation{State Key Laboratory of Low Dimensional Quantum Physics, Department of Physics, Tsinghua University, Beijing 100084, China}
\affiliation{Beijing Academy of Quantum Information Sciences, Beijing 100193, China}
\affiliation{Frontier Science Center for Quantum Information, Beijing 100084, China}
\affiliation{Hefei National Laboratory, Hefei 230088, China}

\author{Tiantian Wang}
\affiliation{Beijing Academy of Quantum Information Sciences, Beijing 100193, China}
\affiliation{Hefei National Laboratory, Hefei 230088, China}

\author{Yunyi Zang}
\affiliation{Beijing Academy of Quantum Information Sciences, Beijing 100193, China}
\affiliation{Hefei National Laboratory, Hefei 230088, China}

\author{Lin Li}
\affiliation{Beijing Academy of Quantum Information Sciences, Beijing 100193, China}

\author{Runan Shang}
\affiliation{Beijing Academy of Quantum Information Sciences, Beijing 100193, China}
\affiliation{Hefei National Laboratory, Hefei 230088, China}

\author{Qi-Kun Xue}
\affiliation{State Key Laboratory of Low Dimensional Quantum Physics, Department of Physics, Tsinghua University, Beijing 100084, China}
\affiliation{Beijing Academy of Quantum Information Sciences, Beijing 100193, China}
\affiliation{Frontier Science Center for Quantum Information, Beijing 100084, China}
\affiliation{Hefei National Laboratory, Hefei 230088, China}
\affiliation{Southern University of Science and Technology, Shenzhen 518055, China}

\author{Ke He}
\email{kehe@tsinghua.edu.cn}
\affiliation{State Key Laboratory of Low Dimensional Quantum Physics, Department of Physics, Tsinghua University, Beijing 100084, China}
\affiliation{Beijing Academy of Quantum Information Sciences, Beijing 100193, China}
\affiliation{Frontier Science Center for Quantum Information, Beijing 100084, China}
\affiliation{Hefei National Laboratory, Hefei 230088, China}

\author{Hao Zhang}
\email{hzquantum@mail.tsinghua.edu.cn}
\affiliation{State Key Laboratory of Low Dimensional Quantum Physics, Department of Physics, Tsinghua University, Beijing 100084, China}
\affiliation{Beijing Academy of Quantum Information Sciences, Beijing 100193, China}
\affiliation{Frontier Science Center for Quantum Information, Beijing 100084, China}


\begin{abstract}

Network structures by semiconductor nanowires hold great promise for advanced quantum devices, especially for applications in topological quantum computing. In this study, we created networks of PbTe nanowires arranged in loop configurations. Using shadow-wall epitaxy, we defined superconducting quantum interference devices (SQUIDs) using the superconductor Pb. These SQUIDs exhibit oscillations in supercurrent upon the scanning of a magnetic field. Most of the oscillations can be fitted assuming a sinusoidal current-phase relation for each Josephson junction. Under certain conditions, the oscillations are found to be skewed, suggesting possible deviation from a sinusoidal behavior. Our results highlight the potential of PbTe nanowires for building complex quantum devices in the form of networks. 

\end{abstract}

\maketitle  

\section{Introduction}

Semiconductor nanowires, as one-dimensional (1D) electron systems, are anticipated to harbor fascinating phases of matter, such as topological superconductivity when coupled to a superconductor \cite{Lutchyn2010, Oreg2010}. While the 1D device configuration greatly facilitates in electron tunability and confinement, it also renders challenges for more intricate device applications. Take Majorana zero modes for instance. While a single 1D wire can serve as a testbed to study relevant experimental signatures \cite{Mourik, Deng2016, Albrecht, Gul2018,  Song2022, WangZhaoyu}, more complex device structures involving nanowire networks are essential for braiding experiments or topological qubits \cite{2010_Alicea, 2013_Hyart, 2016_PRX_milestone, 2016_Fu, 2017_Box_qubit, 2017_PRB_Scalable}. The technique of selective-area-growth (SAG) offers a solution to this scalability issue, enabling the realization of complex network structures through  lithography on demand.

Previous studies on SAG nanowire networks primarily focused on the Aharonov-Bohm effect, where conductance oscillations were observed in a magnetic field due to electron interference \cite{2018_PRL_SAG, Palmstrom_SAG, Pavel_SAG_2019, Roy}. Here, we present the interference of Cooper pairs (supercurrents) in nanowire networks coupled to a superconductor. We choose PbTe as the material, a promising candidate for Majorana nanowires  that has recently emerged \cite{CaoZhanPbTe, Jiangyuying, Erik_PbTe_SAG, PbTe_AB, Fabrizio_PbTe, Zitong, Wenyu_QPC, Yichun, Yuhao, Ruidong, Vlad_PbTe, Wenyu_Disorder, PbTe_In, Yuhao_degeneracy}. Using shadow wall epitaxy of Pb superconducting film, we defined two Josephson junctions (JJs) within the network loop. The resulting dc superconducting quantum interference device (SQUID) exhibit a gate-tunable supercurrent. The supercurrent oscillates in a magnetic field based on which the underlying current-phase relation is discussed. Our results pave the way toward scalable nanowire network devices, with relevance to studies on Majorana zero modes \cite{2016_Fu, 2017_Box_qubit, 2017_PRB_Scalable, NextSteps} and superconductor-semiconductor hybrid qubits \cite{2015_PRL_gatemon, DiCarlo_gatemon,2021_Devoret_Science, Huo_gatemon,2023_NP_Andreev, ZeZhou_PRA, Ramon_perspective}.

\begin{figure*}[htb]
\includegraphics[width=\textwidth]{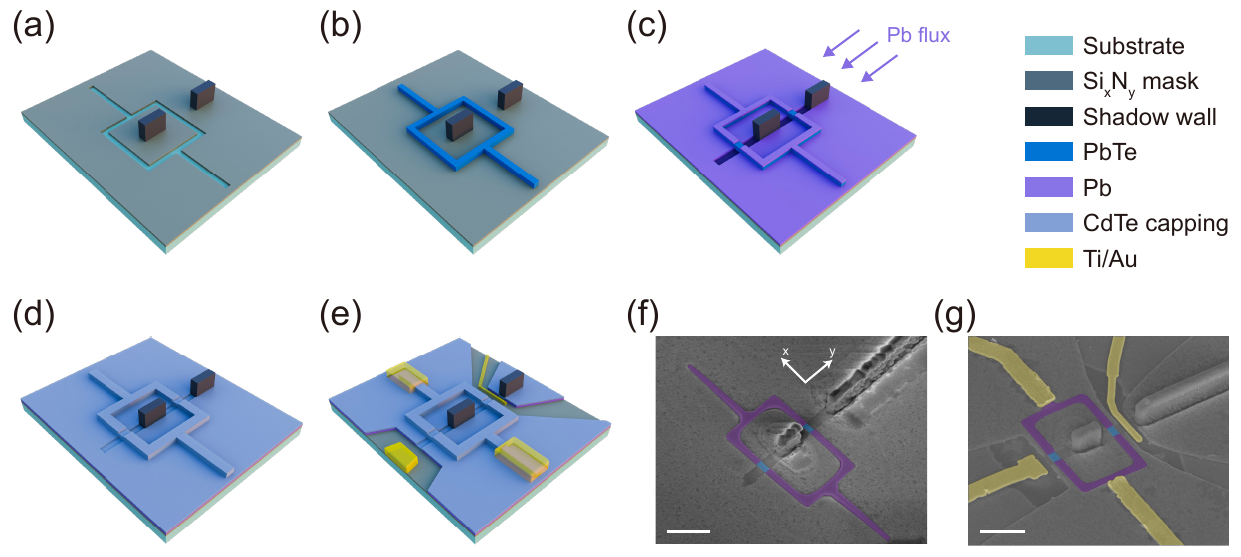}
\centering
\caption{Growth and fabrication of PbTe SQUID networks. (a) A CdTe/Pb$_{1-x}$Eu$_x$Te substrate is covered by a SiN dielectric mask (dark grey) with two Hydrogen SilsesQuioxane shadow walls. A SQUID network pattern is etched on the mask. (b) SAG of PbTe nanowires in the etched trenches. (c) In situ deposition of Pb film. The shadowed region defines Josephson junctions. (d) Growth of CdTe capping. (e) Schematic of a final device with contacts and side gates. (f-g) Tilted SEMs of a typical device corresponding to (d-e), respectively. The scale bar is 1 $\upmu$m. }
\label{fig1}
\end{figure*}

\section{Device growth and fabrication}

Figure 1 illustrates the key steps of the device growth and fabrication. Initially, a CdTe(001) substrate was covered with a thin film of Pb$_{0.92}$Eu$_{0.08}$Te and a CdTe capping layer, grown in a molecular beam epitaxy chamber. Next, a SiN dielectric was deposited on the substrate, and network trenches were defined using reactive ion etching (Fig. 1(a)). Two shadow walls were constructed before these processing steps. Subsequently, PbTe nanowires were selectively grown within the network structure (Fig. 1(b)), followed by the low-temperature deposition of Pb using shadow-wall epitaxy (Fig. 1(c)). The chip was then capped by a thin layer of CdTe (Fig. 1(d)).

Each shadow wall defines a JJ in each ``arm'' of the loop. The two JJs thus form a dc SQUID via the PbTe network. Figure 1(f) is a tilted scanning electron micrograph (SEM) of a representative SQUID. Further details on this growth procedure can be found in Ref. \cite{Wenyu_Disorder}. 

After growth, contacts and side gates were fabricated as shown in Figs. 1(e) and 1(g). To prevent short-circuiting, the regions of Pb film beneath the gates were etched away prior to the gate deposition. Argon plasma etching was conducted in situ before contact deposition to remove the CdTe capping, ensuring ohmic contacts. Devices based on individual PbTe-Pb nanowires grown using the same protocol have exhibited an atomically sharp PbTe-Pb interface, a hard superconducting gap, and gate-tunable supercurrents \cite{Zitong, Yichun}. These findings indicate the high quality of the hybrid nanowires, sufficient for the realization of more intricate network devices.

\section{SQUID oscillations in device A}

Four dc SQUIDs (devices A-D) with varying loop areas were measured, all exhibiting similar behavior. Figure 2(a) presents the SEM of device A (without tilting). The gate voltages applied to the bottom and top gates are denoted as $V_{\text{G1}}$ and $V_{\text{G2}}$, respectively. The measurements were conducted using a standard two-terminal set-up in a dilution refrigerator with a base temperature below 50 mK. Any series resistance, arising from the fridge filters and device contacts, has been subtracted during data processing similar to that in Ref. \cite{Zitong}. 

\begin{figure}[htb]
\includegraphics[width=\columnwidth]{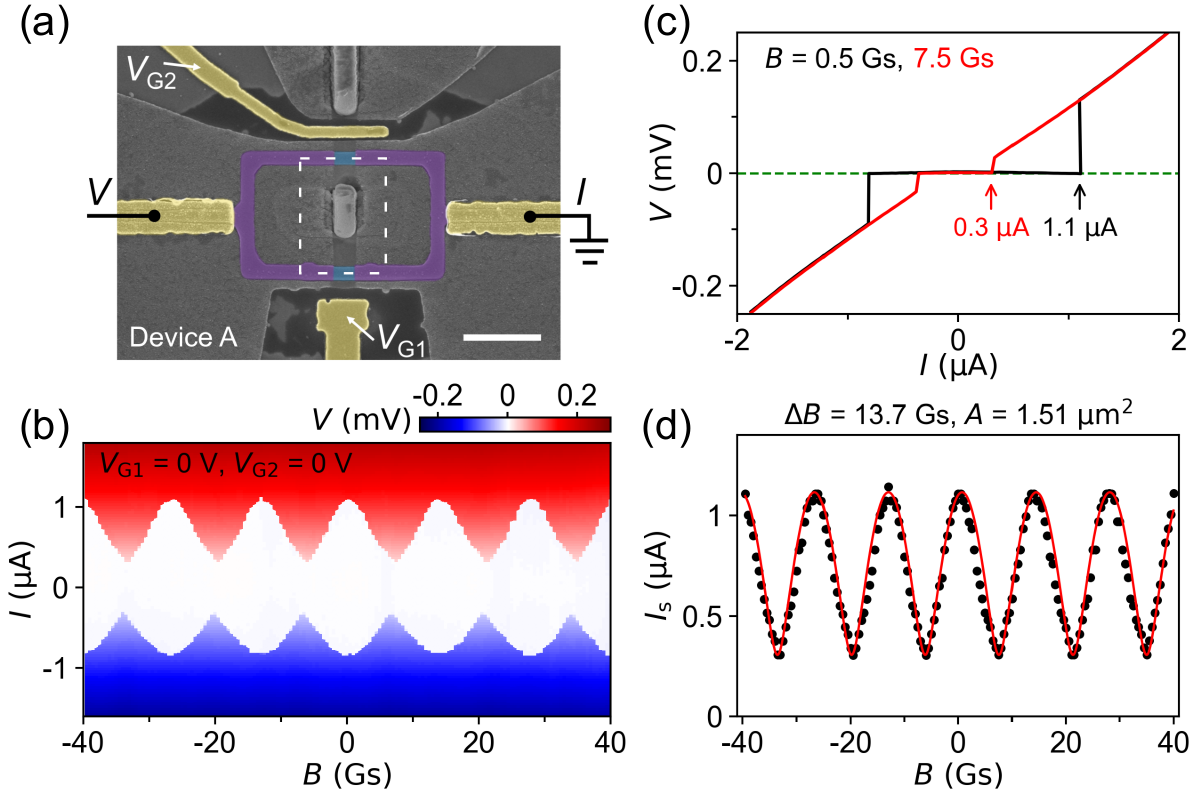}
\centering
\caption{(a) False-colored SEM of device A. The scale bar is 1 $\upmu$m.  (b) $I$-$V$ curve as a function of $B$, showing SQUID oscillations. The sweeping direction of $I$ was from the negative bias to the positive bias.  $V_{\text{G1}} = V_{\text{G2}} =$ 0 V. (c) $I$-$V$ line cuts from (b), at the $B$-values corresponding to the maximum (black) and minimum (red) switching currents. (d) Extracted switching current (the black dots) and a fit (the red curve), assuming a sinusoidal current-phase relation of the two JJs. }
\label{fig2}
\end{figure}

Figure 2(b) shows the voltage drop ($V$) across the SQUID as a function of current ($I$) and magnetic field ($B$) for device A. $B$ was perpendicular to the device substrate throughout the measurement (unless specified). $V_{\text{G1}}$ and $V_{\text{G2}}$ were kept grounded. The white region represents the zero-resistance supercurrent regime. Periodic oscillations of the supercurrent are observed with a period $\Delta B \sim$ 13.7 Gs. Using the formula $\Phi_0=h/2e=\Delta B \times A$, we can convert this period to an effective area ($A$) $\sim$ 1.50 $\upmu$m$^2$, where $\Phi_0$ denotes the flux quantum, $h$ is the Plank constant and $-e$ is the electron charge. The spacing between the top and bottom JJs is $\sim$ 1.5 $\upmu$m. The effective area corresponds to an effective width of $\sim$ 1 $\upmu$m, as indicated by the white dashed box in Fig. 1(a). The box size is larger than the shadowed region but smaller than the network loop, possibly due to the penetration effect of $B$ and flux focusing. Figure 2(c) presents two line cuts with the minimal and maximum switching currents, i.e. 0.3 $\upmu$A and 1.1 $\upmu$A, respectively. Assuming the critical currents of the two JJs are $I_a$ and $I_b$, we then extract $I_a$ = 0.7 $\upmu$A, $I_b$ = 0.4 $\upmu$A, based on $I_a+I_b$ = 1.1 $\upmu$A and $I_a-I_b$ = 0.3 $\upmu$A. It is assumed that the switching current is close to the critical current, as the fridge temperature is much less than the Josephson energy.

In Fig. 2(d), we plot the switching current, $I_s$, extracted from Fig. 2(b). The red curve represents a fit with no fitting parameters (except for a phase offset), assuming a sinusoidal current-phase relation: $i_a = I_a$sin$\phi_a$, $i_b = I_b$sin $\phi_b$. $i_a$ and $i_b$ are the currents passing through the two JJs, and $\phi_a$ and $\phi_b$ are the phase drops over the two JJs. The phases and the magnetic flux through the effective area $\Phi = B \cdot A$ are related by $\phi_a-\phi_b = 2\pi\Phi/\Phi_0$. We have assumed that the contribution from the loop inductance, $LI_{circ}$, is negligible, as the $B$ positions of maximum switching current for positive and negative $I$ biases are nearly identical. Here, $L$ denotes the SQUID inductance and $I_{circ}$ represents the circulating current in the loop. The switching current can then be calculated as $I_s(B)=\sqrt{I_a^2+I_b^2+2I_aI_b\text{cos}(2\pi\Phi/\Phi_0)}$. We plot this formula, based on the extracted $I_a$ and $I_b$, as the red curve in Fig. 2(d), which agrees reasonably well with the experimental data.

\section{Gate dependence of SQUID oscillations in device A}

\begin{figure}[tb]
\includegraphics[width=\columnwidth]{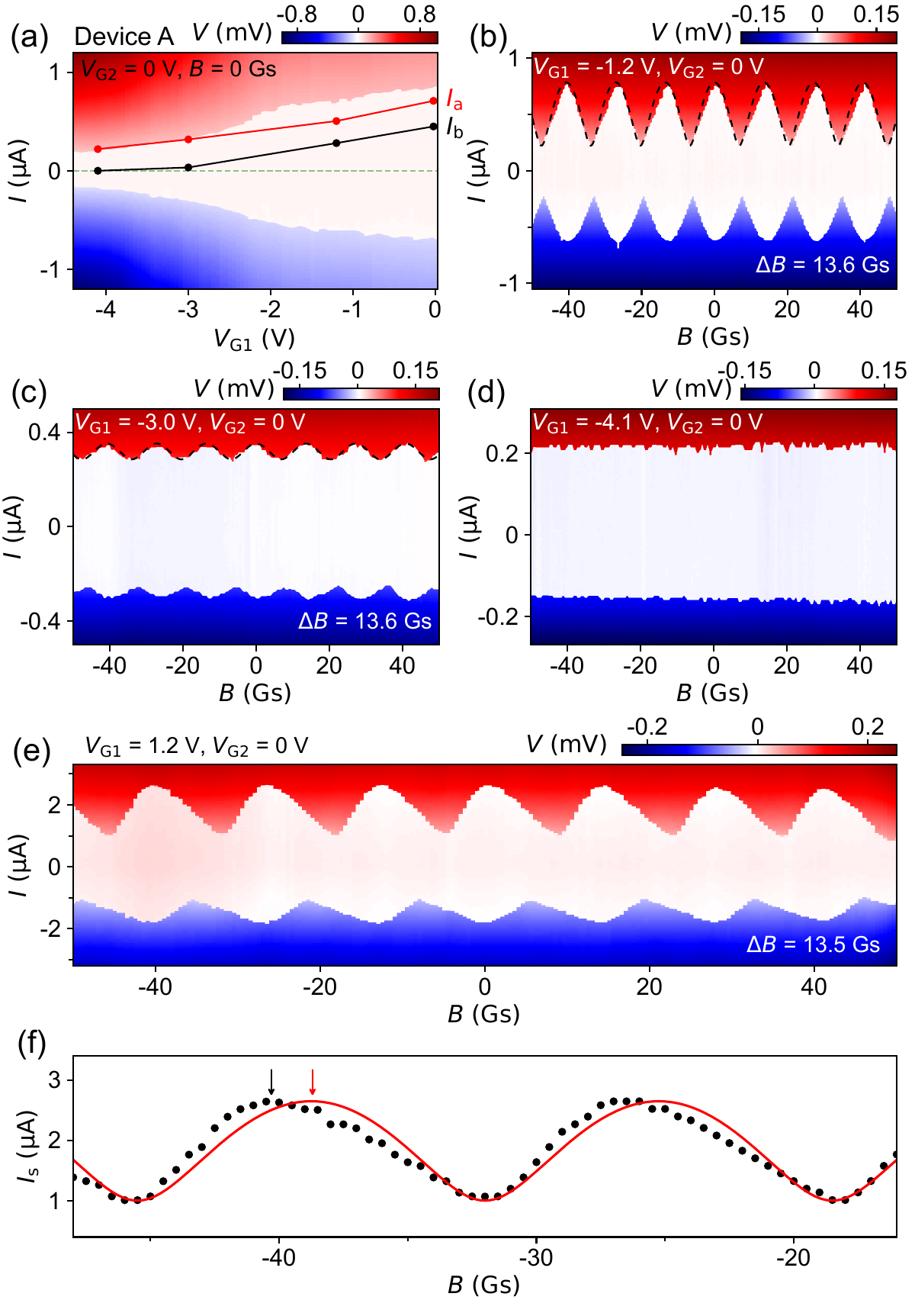}
\centering
\caption{(a) Gate dependence of $I$-$V$ for device A. $B$ = 0 T. The red and black dots are switching currents of the two junctions, extracted based on the SQUID oscillation amplitudes.  (b-d) SQUID measurements of device A at $V_{\text{G1}}$ =  -1.2 V, -3.0 V, and -4.1 V, respectively. $V_{\text{G2}}$ =  0 V. Dashed lines in (b-c) are fittings using the aforementioned method. No oscillations are observed in (d). (e) $V_{\text{G1}}$ = 1.2 V. The oscillation patterns are skewed, deviating from the fit. (f) Extracted $I_s$ (the black dots) from (e) and the fit (the red line), assuming a sinusoidal current-phase relation.    }
\label{fig3}
\end{figure}

Next, we investigate the gate dependence of device A. Figure 3(a) shows the supercurrent as a function of $V_{\text{G1}}$ at $B$ = 0 T, revealing a monotonically decreasing trend in $I_s$. To determine the critical current of each JJ, we conduct the SQUID measurements at four $V_{\text{G1}}$ settings, as illustrated in Fig. 2(b) and Figs. 3(b-d). From the maximum and minimum switching currents of the SQUID oscillations, we infer $I_a$ and $I_b$, and mark them in Fig. 3(a) as red and black dots, respectively. Despite the significant distance between the G1 gate and the top JJ (nearly 2 $\upmu$m), $V_{\text{G1}}$ can still simultaneously modulate the critical currents of both JJs with similar amplitudes. This strong capacitive coupling arises from the global distribution of the PbEuTe substrate, as detailed in Ref. \cite{Wenyu_Disorder} (Fig. S5).

The dashed lines in Figs. 3(b-c) depict fits using the aforementioned method. The fit roughly matches the oscillations with minor deviations, indicating a sinusoidal current-phase relation for both JJs over a considerable gate range. As $I_b$ decreases, the amplitudes of SQUID oscillations diminish while the periods ($\Delta B$) remain unchanged. In Fig. 3(d), no oscillations are observed, suggesting that the supercurrent in one JJ has been turned off.

\begin{figure*}[ht]
\includegraphics[width=0.85\textwidth]{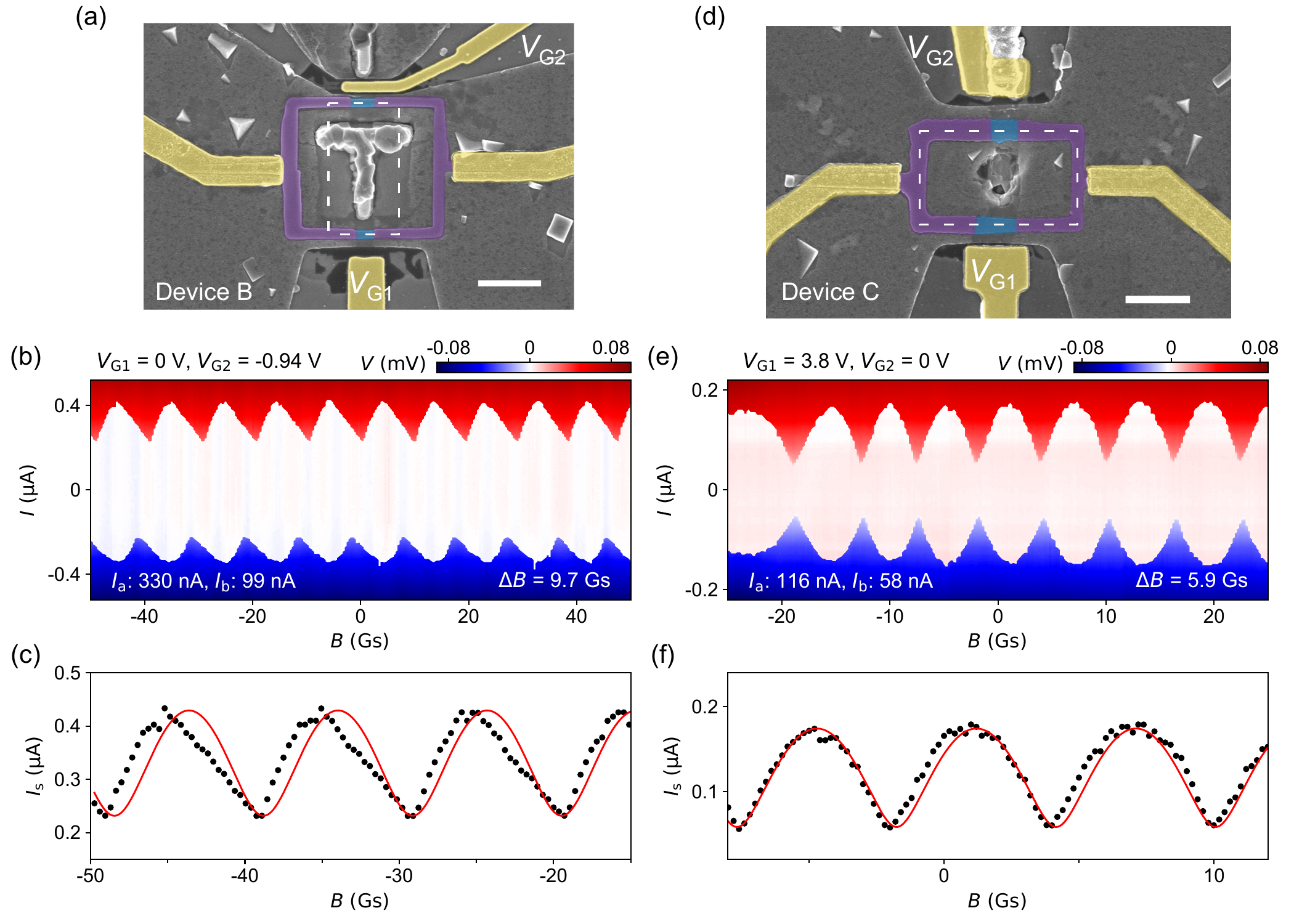}
\centering
\caption{(a) False-colored SEM of device B. The scale bar is 1 $\upmu$m. (b) SQUID measurement of device B at $V_{\text{G1}}$ =  0 V, $V_{\text{G2}}$ =  -0.94 V.  (c) Extracted $I_s$ (the black dots) and a fit (the red line) assuming a sinusoidal current-phase relation. (d) False-colored SEM of device C. The scale bar is 1 $\upmu$m. (e) SQUID measurement of device C at $V_{\text{G1}}$ =  3.8 V, $V_{\text{G2}}$ = 0 V. The oscillation period on the left side is slightly larger than others, possibly due to the instability of the device. (f) Extracted $I_s$ (the black dots) and a fit (the red line). }
\label{fig4}
\end{figure*}

Following this, a skewed oscillation pattern is observed at $V_{\text{G1}}$ = 1.2 V, as illustrated in Fig. 3(e). Figure 3(f) presents the extracted switching currents (depicted as black dots) alongside the fit (represented by the red curve), assuming a sinusoidal current-phase relation of the JJs. For the sake of clarity only two periods are plotted. Notably, the position of maximum $I_s$ (indicated by the black arrow) does not align with the center of the two $B$'s corresponding to the minimum $I_s$ (denoted by the red arrow), indicative of the observed the skewness. Such deviation, also (barely) visible in Fig. 2(b), may imply deviations from a sinusoidal current-phase relation. A non-sinusoidal case is commonly expected for S-N-S JJs exhibiting high junction transparency \cite{1991_PRL_CPR, 2017_Moler_CPR} (S represents superconductor and N denotes normal conductor). While a sinusoidal current-phase relation is typically a valid approximation in scenarios when the junction transmission is small, it may not hold for high transparent junctions. It is essential to note that the current-phase relation cannot be directly discerned from $I_s (B)$ in Fig. 3(f), especially considering that the two JJs are not in the high asymmetry regime. Even in instances of high asymmetry, i.e. $I_a \gg I_b$, the conventional approach of using $I_s(B)$ as an approximation of $i_b(\phi_b)$ \cite{2007_CPR, Srijit_CPR, TI_CPR} has its inherent limitations \cite{CPR_WTe, CPR_limitation}.  For an overview of SQUID oscillations in device A at various $V_{\text{G}}$ settings, we refer to Fig. S1 in the Supplementary Material.

\section{SQUID oscillations in  additional devices}

Figure 4 presents two additional devices  with varying network sizes (loop areas), showcasing analogous behavior. In Fig. 4(a), the SEM of device B reveals triangle debris outside loop, attributed to PbTe parasitic growth, as the selectivity of this particular growth is slightly inferior. The shadow wall adopted a T shape, resulting in a larger shadowed region inside the loop compared to device A. The loop size is also larger. Consequently, the period of the SQUID oscillations in Fig. 4(b), denoted as $\Delta B \sim$ 9.7 Gs, is smaller than that observed in device A. This period can be converted to an effective area of 2.12 $\upmu$m$^2$, as indicated by the white dashed box in Fig. 4(a). The extracted switching current (depicted as black dots in Fig. 4(c)) exhibits deviation from the fit (illustrated by the red curve), similar to the behavior observed in Fig. 3(f). The skewed portion has a shoulder/kink structure near $I_s$ of 0.35 $\upmu$A, reminiscent of the current-phase relation of Andreev bound states in the presence of charging energy \cite{2019_PRB_CPR}.

Figure 4(d) shows the SEM of a third device. The shadow wall inside the loop detached from the substrate chip during the device fabrication process. The effective area, determined based on the period of the SQUID oscillations in Fig. 4(e), is $A \sim$ 3.5 $\upmu$m$^2$, as indicated by the white dashed box. Unlike devices A and B, the box size matches the network loop of device C. The underlying mechanisms governing flux focusing and $B$ penetration for each specific device remain unclear. Nevertheless, the oscillations in device C are roughly consistent with a sinusoidal current-phase relation, as shown in Fig. 4(f). For additional SQUID oscillations of devices B and C at various gate settings, we refer to Figs. S2 and S3 in the Supplementary Material. Furthermore, Fig. S4 depicts a fourth device, which also exhibits gate-tunable SQUID oscillations.

\section{Conclusion and Outlook}

In summary, we have realized dc SQUIDs using selective-area-grown PbTe-Pb nanowire networks. These networks exhibit tunable supercurrents and SQUID oscillations in response to a magnetic field. The underlying current-phase relations have been explored. The PbTe-Pb hybrid nanowires hold promise for advancing topological quantum computation in the quest for Majorana zero modes. While transport studies on individual nanowires offer valuable insights, they are insufficient for definitive proving the existence of Majoranas. Braiding experiments, necessitating intricate network architectures, are imperative for conclusive validation. Our results represent a stride toward scalable network devices, heralding prospects for more intricate geometries such as topological qubits.

Raw data and processing codes within this paper are available at https://doi.org/10.5281/zenodo.10952569

\section{Acknowledgment}

This work is supported by National Natural Science Foundation of China (92065206) and the Innovation Program for Quantum Science and Technology (2021ZD0302400).

\bibliography{mybibfile}

\newpage

\onecolumngrid

\newpage


\section*{Supplementary material}

# Supplemental Material for "SQUID oscillations in PbTe nanowire networks"

Yichun Gao,[1, *] Wenyu Song,[1, *] Zehao Yu,[1, *] Shuai Yang,[1] Yuhao Wang,[1] Ruidong Li,[1] Fangting Chen,[1] Zuhan Geng,[1] Lining Yang,[1] Jiaye Xu,[1] Zhaoyu Wang,[1] Zonglin Li,[1] Shan Zhang,[1] Xiao Feng,[1, 2, 3, 4] Tiantian Wang,[2, 4] Yunyi Zang,[2, 4] Lin Li,[2] Runan Shang,[2, 4] Qi-Kun Xue,[1, 2, 3, 4, 5] Ke He,[1, 2, 3, 4, †] and Hao Zhang[1, 2, 3, ‡]

[1]*State Key Laboratory of Low Dimensional Quantum Physics, Department of Physics, Tsinghua University, Beijing 100084, China*
[2]*Beijing Academy of Quantum Information Sciences, Beijing 100193, China*
[3]*Frontier Science Center for Quantum Information, Beijing 100084, China*
[4]*Hefei National Laboratory, Hefei 230088, China*
[5]*Southern University of Science and Technology, Shenzhen 518055, China*

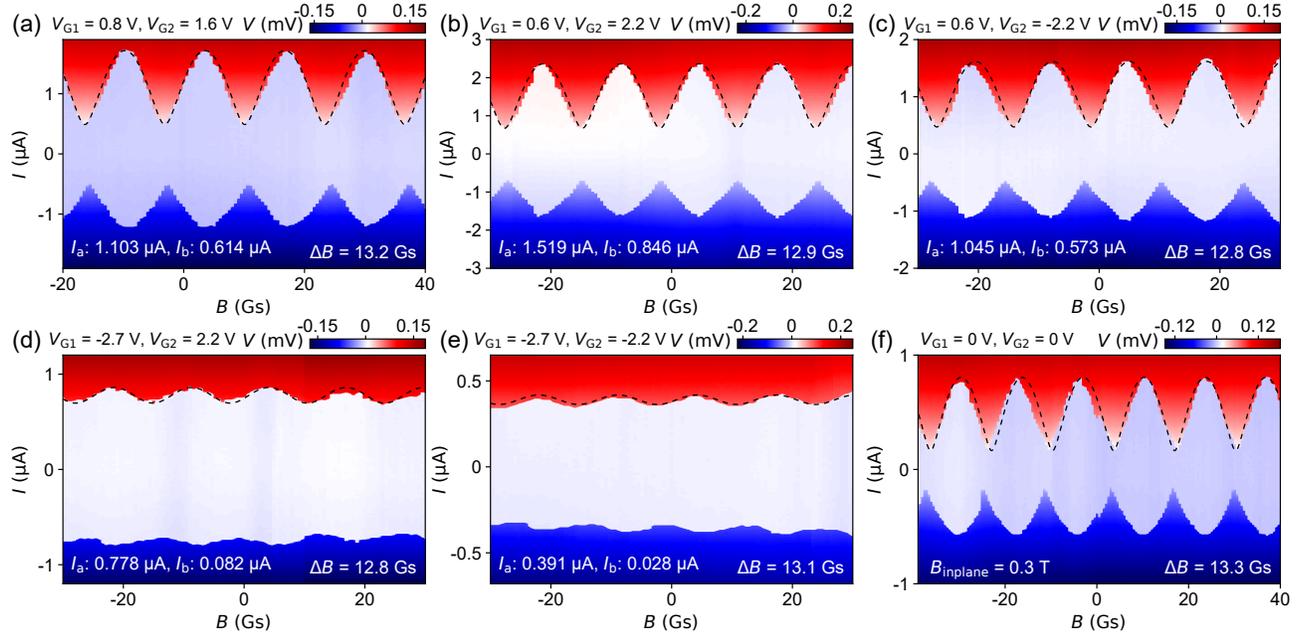

FIG. S1. Additional SQUID measurements for device A. Gate voltages, periods $\Delta B$, and values of $I_a$ and $I_b$ are labeled in each panel. The dashed lines are fits. For panel (f), an in-plane magnetic field of 0.3 T was applied.

---

* equal contribution
† kehe@tsinghua.edu.cn
‡ hzquantum@mail.tsinghua.edu.cn

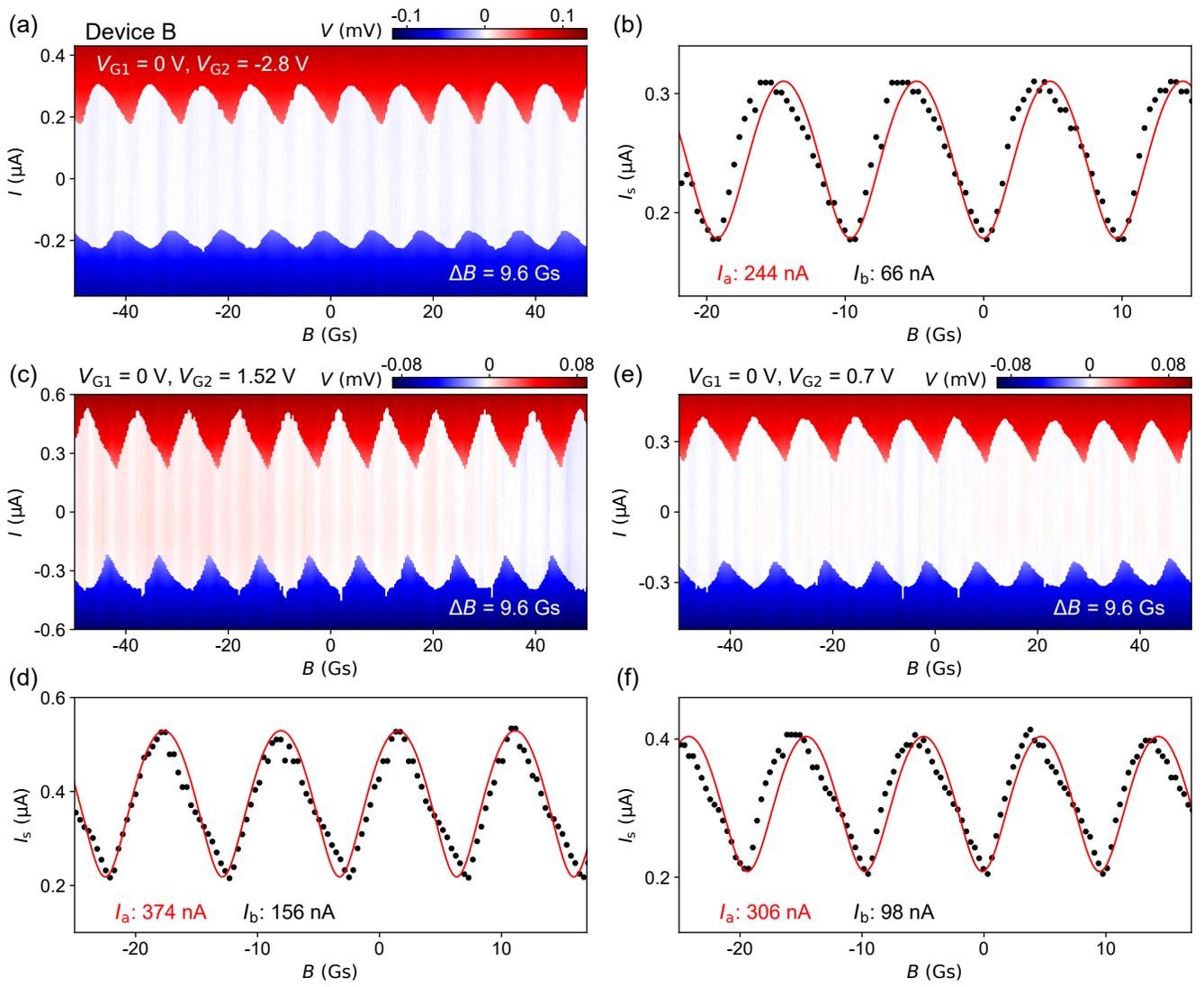

FIG. S2. Three additional SQUID measurements for device B and the extracted switching currents. A kink feature similar to that of Fig. 4(c) is observable.

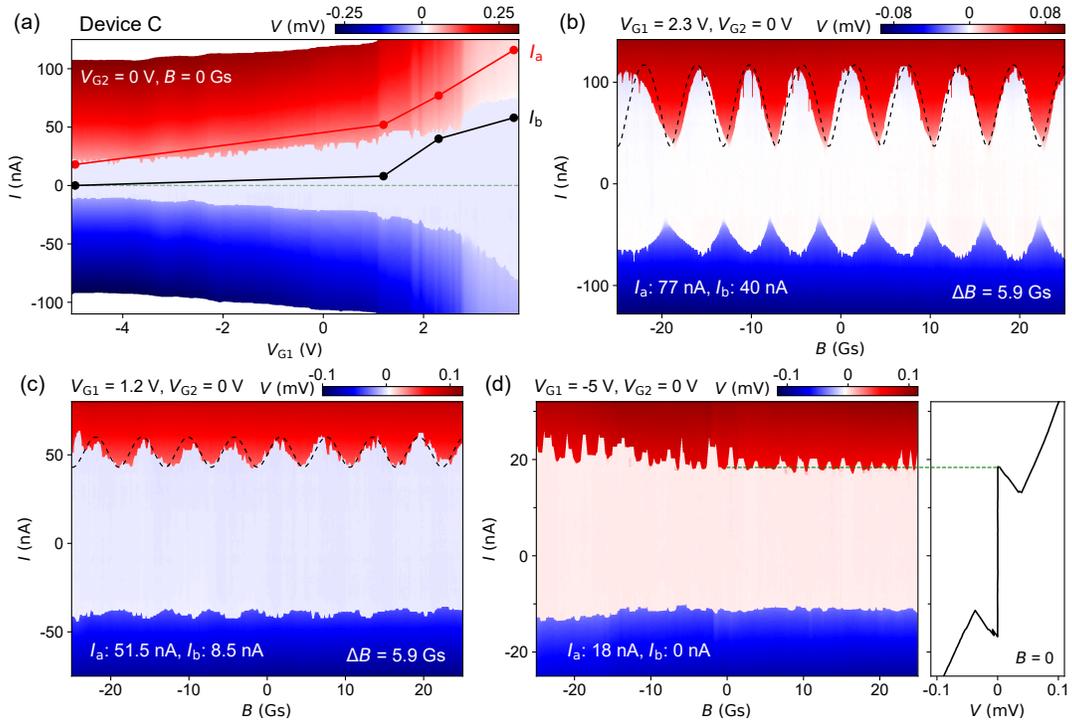

FIG. S3. (a) Gate dependence of the supercurrent in device C. The scatter points are $I_a$ and $I_b$, extracted from the SQUID oscillations in (b-d) and Fig. 4(e). (b-d) SQUID measurements at three different $V_{G1}$'s. The oscillation amplitudes decreases as $V_{G1}$ goes more negative. No SQUID oscillations are observed in (d). Right panel of (d) shows a line cut of the $I$-$V$ curve. Near the switching region, each $I$ corresponds to two value of $V$. We plot the color map so that the positive bias branch can accurately reflect the switching current.

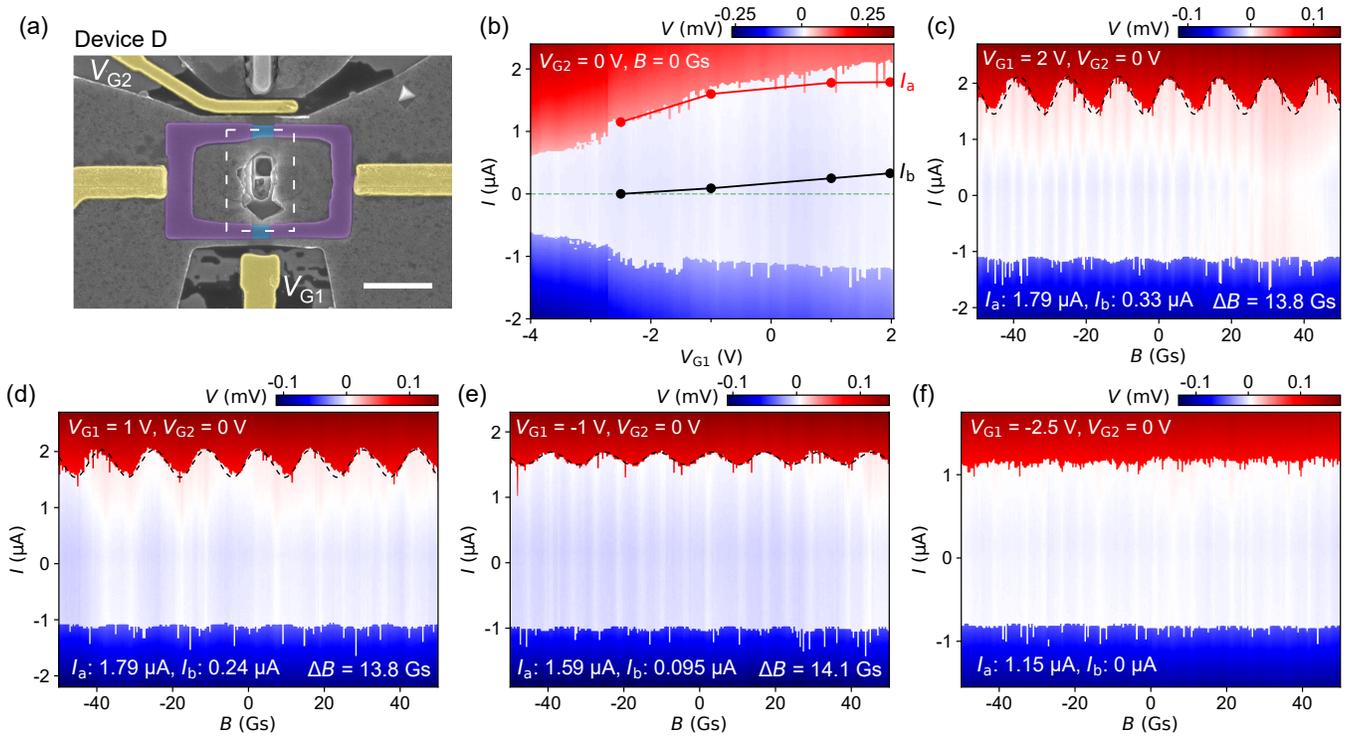

FIG. S4. SQUID measurements of device D. (a) False-colored SEM. The scale bar is 1 μm. The dashed box is the effective loop area, estimated based on the oscillation periods. (b) Gate dependence of the supercurrent and extracted critical currents of the two JJs. (c-f) SQUID measurements at four gate settings.



\end{document}